\documentclass[journal=mamobx,manuscript=article]{achemso}
\setkeys{acs}{articletitle = true}
\setkeys{acs}{etalmode=truncate, maxauthors=100}
\usepackage{chemformula} 
\usepackage[T1]{fontenc} 

\usepackage[normalem]{ulem}
\newcommand\DEL[1]{\textcolor{red}{\sout{}}}     
\author{Christophe Brouzet}
\affiliation[KTH Royal Institute of Technology]
{Linn{\'e} FLOW Centre, KTH Mechanics, KTH Royal Institute of Technology, Stockholm SE-100 44, Sweden}
\alsoaffiliation[KTH Royal Institute of Technology]
{Wallenberg Wood Science Center, KTH Royal Institute of Technology, Stockholm SE-100 44, Sweden}

\author{Nitesh Mittal}
\affiliation[KTH Royal Institute of Technology]
{Linn{\'e} FLOW Centre, KTH Mechanics, KTH Royal Institute of Technology, Stockholm SE-100 44, Sweden}
\alsoaffiliation[KTH Royal Institute of Technology]
{Wallenberg Wood Science Center, KTH Royal Institute of Technology, Stockholm SE-100 44, Sweden}

\author{Fredrik Lundell}
\affiliation[KTH Royal Institute of Technology]
{Linn{\'e} FLOW Centre, KTH Mechanics, KTH Royal Institute of Technology, Stockholm SE-100 44, Sweden}
\alsoaffiliation[KTH Royal Institute of Technology]
{Wallenberg Wood Science Center, KTH Royal Institute of Technology, Stockholm SE-100 44, Sweden}
\email{frlu@kth.se}

\author{L. Daniel S{\"o}derberg}
\affiliation[KTH Royal Institute of Technology]
{Linn{\'e} FLOW Centre, KTH Mechanics, KTH Royal Institute of Technology, Stockholm SE-100 44, Sweden}
\alsoaffiliation[KTH Royal Institute of Technology]
{Wallenberg Wood Science Center, KTH Royal Institute of Technology, Stockholm SE-100 44, Sweden}

\title[Characterizing the Orientational and Network Dynamics of Polydisperse Nanofibres at the Nanoscale]
  {Characterizing the Orientational and Network Dynamics of Polydisperse Nanofibres at the Nanoscale}

\abbreviations{CNF: cellulose nanofibres, CNC: cellulose nanocrystals}

\begin{document}

\begin{tocentry}

%
%
%

\includegraphics[width=8.3cm,height=3.5cm]{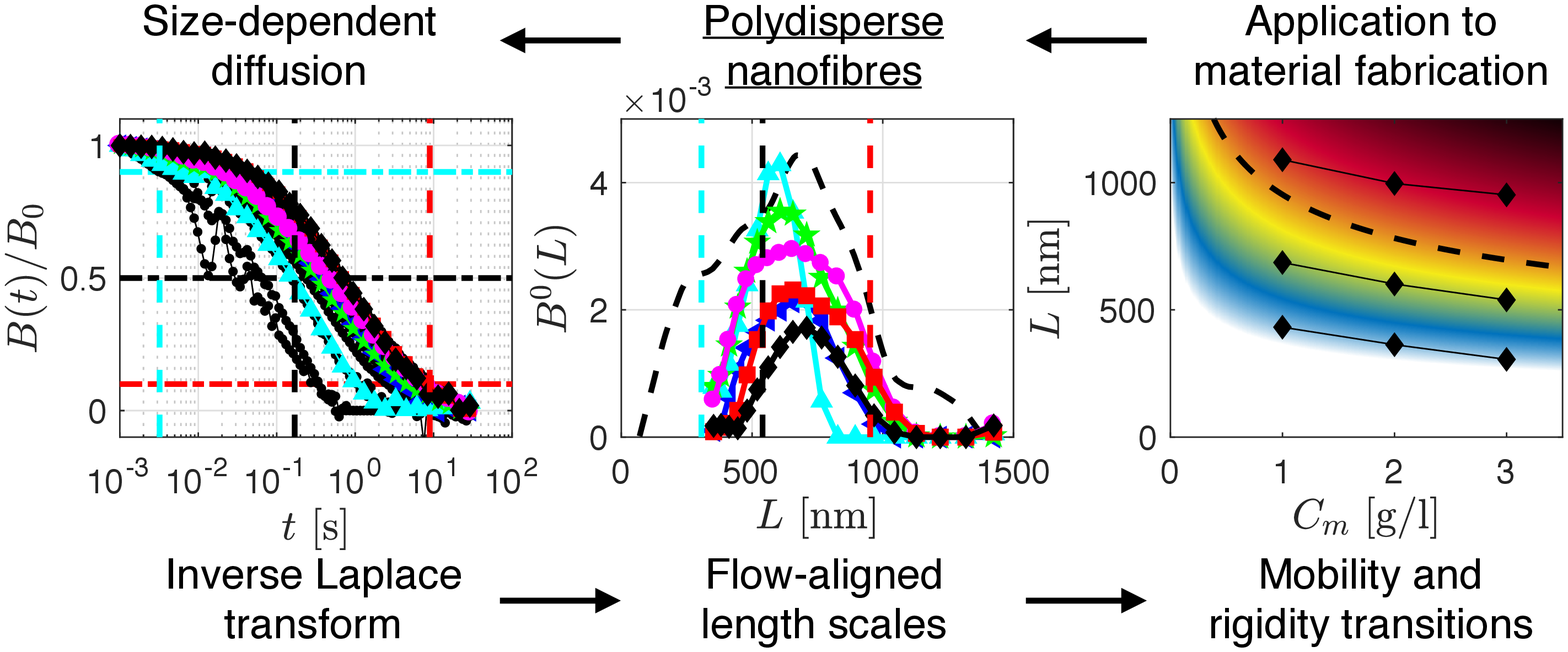}

\end{tocentry}

\begin{abstract}

Polydisperse fibre networks are the basis of many natural and man-made architectures, ranging from high-performance bio-based materials to components of living cells and tissues. The formation and persistence of such networks are given by fibre properties such as length and stiffness as well as the number density and fibre-fibre interactions. Studies of fibre network behavior, such as connectivity or rigidity thresholds, typically assume fixed fibre length and isotropic fibre orientation distributions, specifically for nanoscale fibres where the methods providing time-resolved measurements are limited. Using birefringence measurements in a microfluidic flow-focusing channel combined with a flow-stop procedure, we here propose a methodology allowing investigations of length dependent rotational dynamics of nanoscale polydisperse fibre suspensions, including effects of non-isotropic orientation distributions. Transition from rotational mobility to rigidity at entanglement thresholds is specifically addressed for a number of nanocellulose suspensions, which are used as model nanofibre systems. The results show that the proposed method allows characterization of the subtle interplay between Brownian diffusion and nanoparticle alignment on network dynamics.

\end{abstract}



\section{Introduction}

Fibre networks are ubiquitous in nature ranging from arterioles inside the body, extracellular matrices generated by the cells to the man-made materials such as papers, textiles, polymer solar cells, and injectable hydrogels~\cite{BravermanFonferko1982,Hartgerinketal2001,Picu2011,Huetal2011,Zhangetal2012,Yangetal2012,Franekeretal2015,Mittaletal2017}. Mechanical and dynamical performances of these networks are given by the physical and chemical properties of individual fibres, the fibre-fibre interactions and nature of bonds as well as the disordered network architecture. Investigating how fibre network architecture at the nanoscale affects mechanical performance, and serves as a basis for improved fabrication and design strategies, has became crucial in the development of novel engineered materials. A mechanistic understanding of the assembly methods and of the architecture at the nanoscale could lead to the generation of high-performance engineered constructs. Current imaging techniques such as electron microscopy that are typical to get information on the fibre architectures, are limited to dry systems~\cite{Stallardetal2018,Mittaletal2018}. Other imaging methods such as dynamic light scattering~\cite{Hassanetal2015}, holographic video microscopy~\cite{CheongGrier2010}, and confocal microscopy~\cite{Edmondetal2012} that have been extensively used to characterize wet colloidal systems, are mostly limited to monodisperse spherical particles or dilute suspensions. Hence, these techniques are unable to thoroughly characterize dynamic colloidal networks of high-aspect ratio nanoparticles such as nanofibres or nanorods, specifically when these particles have a wide range of length distributions.

Macroscale mechanical performance given by the network structure~\cite{Picu2011,Mulleretal2000,Rothetal2003} depends on the interplay between nanoscale thermal fluctuations, causing rotational and translational diffusion, and entanglement of nanofibres in a connected network. Hydrodynamic assembly is a promising way to control nanofibre motion during processing~\cite{NatureCom2014,Kamadaetal2017,Mittaletal2018}, where shear and extensional flows are used to control nanofibre alignment~\cite{Jeffery1922} in competition with Brownian diffusion and under the influence of network connectivity. Despite a vast literature on Brownian motion of macromolecules~\cite{DoiEdwards1986,TracyPecora1992}, our understanding of the thermal and orientational dynamics of nanoparticles in a crowded environment is largely incomplete~\cite{Fakhrietal2010}. This is due to the lack of \textit{in-situ} characterization techniques that are able to trace in real-time the complex dynamics of nanoparticles, specifically under dynamic flow conditions.

With suspensions of polydisperse nanofibres, as is the case of most of the nanofibres of biological origin, the orientational dynamics largely depends on the nanoparticle size~\cite{MarrucciGrizzuti1983,MarrucciGrizzuti1984,KeepPecora1985,Rogersetal2005,FlowStopLetter2018}. Polydispersity has a significant influence on the dynamics of nanofibre networks as well as on the mechanical properties of the nanostructured macroscopic materials. A broad length distribution further challenges our understanding on the physical description of the nanoscale entangled network as nanofibres experience multiple entanglement regimes depending on their length and as polydispersity increases the mean entanglement significantly~\cite{KrophollerSampson2001,Huberetal2003}. 
For man-made materials, the nanofibre assembly under the influence of Brownian motion can be controlled by tuning the concentration and nanofibre dimensions in a way such that entanglement restrains Brownian motion without impeding flow-alignment~\cite{Mittaletal2018}.

In this paper we present a methodology to characterize the length-dependent rotational dynamics of polydisperse nanofibre suspensions, including the effects from non-isotropic orientation distributions,
and the entanglement in a connected network as a function of nanofibre length. The experimental set-up consists of a flow-focusing channel~\cite{RotDiffTomas}. A birefringence dependent flow-stop methodology recently introduced~\cite{RotDiffTomas,FlowStopLetter2018} is then applied and extended to well characterized suspensions (in dry state) of cellulose nanofibres (CNF) and cellulose nanocrystals (CNC), with different length distributions and at different concentrations. Obtained results demonstrate that \textit{in-situ} dynamics of polydisperse nanofibre systems can be well characterized and is highly dependent on nanoparticle length distributions and on entanglement at increased concentrations. We further discuss the implications of current findings in the context of the performance of nanostructured materials fabricated \textit{via} hydrodynamic approaches.

\section{Experimental section}

A total of six different nanofibre suspensions are used in this study with CNC as stiff nanorods having a narrow length distribution and with CNF as rigid nanofibres having different surface charge densities, which in turn lead to nanofibres with different length distributions. Each sample has been used at three different concentrations. This allows us to get a thorough understanding of the network dynamics of nanoparticles systems with different length distributions, entanglement and surface charge densities.

\subsection{Nanocellulose suspensions}

The preparation of the nanocellulose suspensions is detailed in the Supporting Information. The concentrations of the suspensions were determined using evaporation and  gravimetric analysis. The surface charges of the nanorods ($103~\mu$eq/g) and nanofibres ($380$, $550$, $820$, $980$ and $1360~\mu$eq/g) were determined by polyelectrolyte titration using a Stabino particle charge mapping equipment with streaming potential measurements (ParticleMetrix, Germany).

The length distributions were measured using transmission electron microscopy (TEM) (JEOL JEM-1400 TEM) with similar protocols used in earlier studies~\cite{Gengetal2018,Mittaletal2018,FlowStopLetter2018}. Images were acquired systematically and randomly to avoid bias (see Figure~S$1$). For each sample, around $200$~nanorods and nanofibres have been used for the length distribution.

The diameter of the nanorods were measured~\cite{FlowStopLetter2018} directly from the high resolution TEM images (Figure~S$2$). The diameter of the nanofibres were measured using atomic force microscopy (AFM)~\cite{Mittaletal2018,FlowStopLetter2018} and small angle X-ray scattering (SAXS)~\cite{Gengetal2018}. For AFM measurements (MultiMode~$8$, Bruker, Santa Barbara, CA, USA), all nanofibres have been assumed to be cylindrical. Therefore the height measured with the AFM corresponds directly to the diameter of the nanofibres. For SAXS measurements ($12$-ID-B beamline, Advanced Photon Source, Argonne National Laboratory, USA), the data were fitted to obtain the typical sizes of the nanofibre cross-section using a ribbon model~\cite{Suetal2014}. For simplification, we consider here that the nanofibres have a circular cross-section and the diameter is therefore obtained from the SAXS data by conversion from ribbon-like to circular cross sections of the same area. The diameters measured by AFM and by SAXS are in good agreement.

\subsection{Experimental setup}

The experimental set-up consists of a flow-focusing channel~\cite{Nunesetal2013,RotDiffTomas}. The channel has a cross shape with four branches, each branch having a square cross-section of side $h=1$~mm. The nanocellulose suspension flows in one branch towards the centre of the cross while two distilled water sheath flows enter in the channel through the two neighboring perpendicular branches. They focus the suspension into a thread flowing away from the centre of the cross in the last branch. This gives rise to an extensional flow, aligning the nanoparticles along the flow direction~\cite{Kiriyaetal2012,NatureCom2014}. The relative orientation of the nanoparticles in the flow is measured through the birefringence properties of the suspension~\cite{Chowetal1985,Rogersetal2005,Rogersetal2005b,RotDiffTomas,FlowStopLetter2018}. The channel is placed between two crossed polarizers and a high-speed camera (SpeedSense M, Dantec Dynamics) collects a red laser light passed through the whole system. As the measured birefringence also depends on the laser intensity and the concentration of the suspension, the data are normalised with a reference laser intensity and the concentration dependency is corrected accordingly. It is therefore possible to compare the birefringence values obtained for the different suspensions. In addition, each branch of the channel is equipped with a solenoid-driven slider valve (MTV-3SL Series, Takasago Electric, Inc.)~\cite{RotDiffTomas}. This allows to rapidly stop the flow once it has reached a steady state and to observe the orientation relaxation, i.e. the decay of the birefringence, caused by the dealignment of the nanorods and nanofibres with the Brownian motion. The decay dynamics is sampled on more than $3$~decades for the nanorods and $4$~decades for nanofibres~\cite{FlowStopLetter2018}.

\subsection{Inverse Laplace transform}

Birefringence decays for polydisperse systems exhibits multiple time scales~\cite{Chowetal1985,Rogersetal2005,RotDiffTomas,FlowStopLetter2018} as the diffusion coefficient strongly depends on the nanoparticle length $L$. Therefore, deconvoluting the birefringence decays $B(t)$ allows to obtain the contribution, i.e. the relative orientation, $B^0(L)$ of the nanofibres of length~$L$ to the total birefringence signal $B_0$ before the flow is stopped~\cite{Rogersetal2005,FlowStopLetter2018}. The deconvolution is performed using an Inverse Laplace transform (see details in the Supporting Information) combined with a rotational diffusion coefficient for a polydisperse system~\cite{MarrucciGrizzuti1983,MarrucciGrizzuti1984}
\begin{equation}
D_r(L)\approx\frac{\beta k_B TL_*^4}{\eta L^7}.\label{eq:polydisperse}
\end{equation}
Here $\beta$~is a numerical factor, $k_B$~the Boltzmann constant, $T$~the temperature, $L_*$~the entanglement length (see Supporting Information) and $\eta$~the solvent viscosity. $\beta$ has been predicted to be of order $1-10$~\cite{Pecora1985,Chowetal1985,KeepPecora1985,Teraokaetal1985,Rogersetal2005} but its magnitude has been found in the range of $10^3-10^4$ by several experiments~\cite{Pecora1985,Chowetal1985,Rogersetal2005,FlowStopLetter2018} and refined theories~\cite{KeepPecora1985,Teraokaetal1985}.

\section{Results and discussion}

\subsection{Characteristics of the nanocellulose suspensions}

\begin{table}[b!]
  \caption{Characteristics of the different suspensions used in this manuscript.}
  \label{tbl:samples}
  \begin{tabular}{lccccc}
    \hline
    Suspension  & $C_m$~[g/l] & $d$~[nm] & $\langle L \rangle$~[nm] & $\widehat{L}$~[nm] & $\sigma$~[nm] \\
    \hline
    CNF-$380$   & $1-2-3$ & $3.62$ & $633$ & $670$ & $720$ \\
    CNF-$550$ & $1-2-3$ & $3.62$ & $600$ & $550$ & $670$ \\
    CNF-$820$  & $1-2-3$ & $3.02$ & $689$ & $770$ & $710$ \\
    CNF-$980$ & $1-2-3$ & $3.02$ & $543$ & $420$ & $550$ \\
    CNF-$1360$ & $1-2-3$ & $2.95$ & $426$ & $390$ & $380$ \\
    CNC & $11-20-41$ & $15$ & $151$ & $130$ & $125$ \\
    \hline
  \end{tabular}
\end{table}

Characteristics of the $18$~different suspensions used in this study are depicted in Table~\ref{tbl:samples}. The nanofibre samples are named using a suffix which represents the surface charges in $\mu$eq/g. For each surface charge density including nanorods, suspensions with three different concentrations have been used ($11$, $20$ and $41$~g/l for the nanorods and $1$, $2$ and $3$~g/l for the nanofibres). These concentrations in particular have been chosen as nanocellulose suspensions lead to a rigid volume-spanning arrested state~\cite{Mittaletal2018,Nordenstrometal2017} (a gel) at higher concentrations and also because the birefringent signal-over-noise ratio is too low at lower concentrations. Note that the transition range from free flowing suspension to an arrested state are different for nanorods and nanofibres due to the large variations in their aspect ratios~\cite{Nordenstrometal2017}. The diameter of nanoparticles for each sample is almost constant and is around $15$~nm for the nanorods and $3$~nm for the nanofibres (see Experimental Section for details).

\begin{figure*}[t!]
\begin{center}
  \includegraphics[width=1\linewidth,clip=]{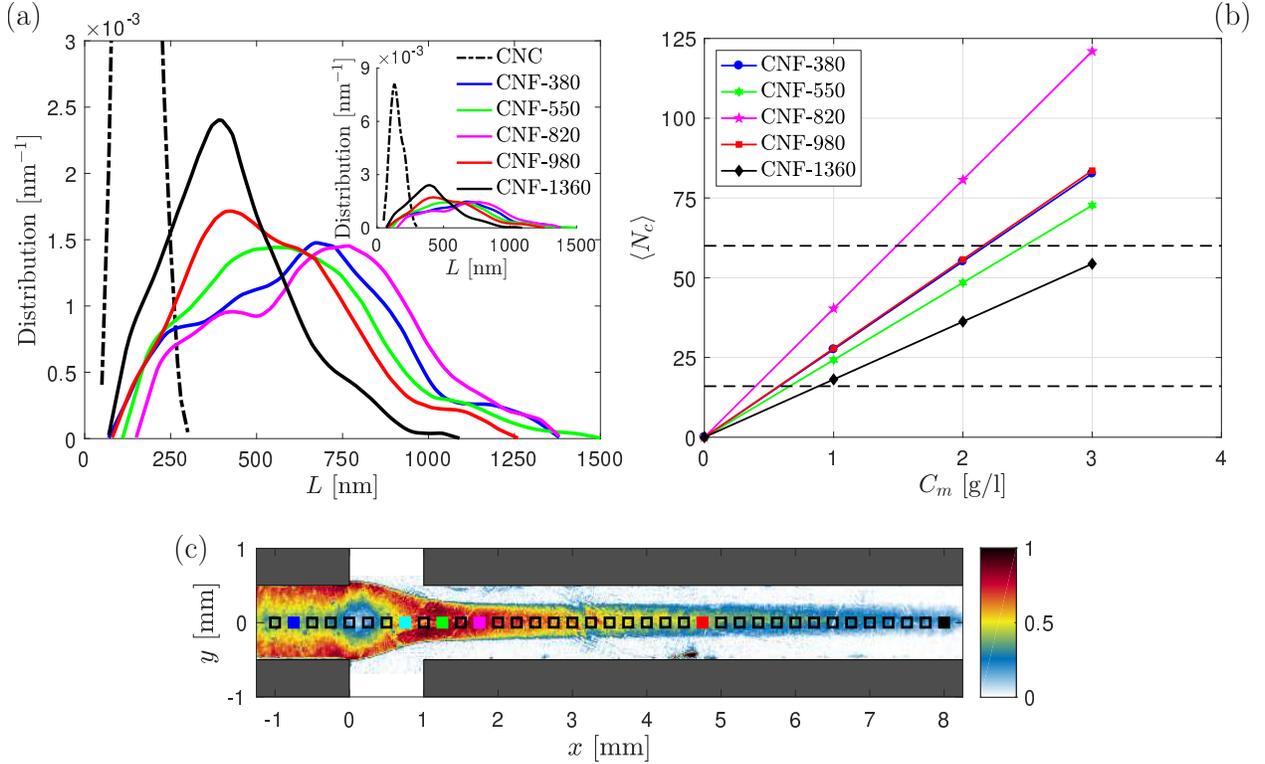}
   \caption{Samples used and experimental set-up. (a) Length distributions of the nanorod (dashed dotted line) and nanofibre (solid line with different colours) suspensions. The insets show the length distributions with a different $y$-coordinate axis to make them all visible. (b)~Mean crowding number $\langle N_c \rangle$ for the different nanofibre suspensions as a function of mass concentration $C_m$. The two horizontal dashed lines represent the connectivity ($\langle N_c \rangle=16$) and rigidity ($\langle N_c \rangle=60$) thresholds. (c) Typical birefringence signal for the CNF-$1360$ suspension. The signal is normalized by its maximum value in the channel. The squares represent the different areas where the birefringence decay is averaged in space and analysed.}\label{fig:setup}
\end{center}
\end{figure*}

The nanoparticle length distributions are plotted in Figure~\ref{fig:setup}(a): the nanorods (dashed dotted line) are almost monodisperse while the nanofibres (coloured solid lines) are polydisperse with different length distributions. These distributions are characterised by their average length $\langle L \rangle$, their most probable length $\widehat{L}$ and their full width at half maximum $\sigma$. These parameters are gathered in Table~\ref{tbl:samples} for all the samples. The three nanofibre samples with the lowest charges (from $380$ to $820$~$\mu$eq/g) have wide length distributions and large lengths of nanofibres while the two nanofibre samples with the highest surface charges ($980$ and $1360$~$\mu$eq/g) show narrower length distributions and shorter lengths of nanofibres as the surface charge density increases~\cite{Gengetal2018}. This suggests a strong coupling between surface charge density and length distributions of nanofibres obtained after fibrillation of pulp fibres, as also observed in earlier studies~\cite{Mittaletal2018,Gengetal2018}.

The crowding number $N_c$ is used to parametrize the entanglement within the network. This dimensionless number is defined as the average number of nanofibres in a sphere with a diameter equal to the length of the fibres~\cite{KerekesSchell1992,Kerekes2006}
\begin{equation}
N_c=\frac{2}{3}\phi \left(\frac{L}{d}\right)^2.
\end{equation}
Here, $L$ is the nanofibre length, $d$ the diameter, and $\phi$ the volume fraction in the suspension. The crowding number allows us to define three different concentration regimes~\cite{Kerekes2006}. For $N_c<16$, the nanofibres are in the dilute regime, with no physical contact between them. In the semi-concentrated intermediate regime where $16<N_c<60$, the nanofibres have significant hydrodynamic and contact interactions but are not immobilized. For suspensions with $N_c>60$, the nanofibres are in a connected network where free motion is hindered. The two thresholds, $N_c=16$ and $N_c=60$, are respectively called connectivity and rigidity thresholds~\cite{Kerekes2006,Celzardetal2009}. In the literature of polymer-based rigid rods~\cite{DoiEdwards1986,KeepPecora1985}, the crowding number is defined as the number of nanofibres in a cube of side $L$. This creates a small discrepancy between the two definitions, with a factor about one half. However, similar concentration regimes can be defined. 

In the case of a network with a nanofibre length distribution, polydispersity gives rise to a distribution of crowding numbers. Different approaches have been suggested to estimate this distribution and the mean crowding number $\langle N_c \rangle$, for a log-normal~\cite{KrophollerSampson2001} or a general~\cite{Huberetal2003} length distribution. In all cases, polydispersity increases the mean crowding number, and therefore the mean entanglement, significantly. The mean crowding numbers, obtained by following the procedure given by Huber \latin{et al.}~\cite{Huberetal2003} for a general length distribution, are shown in Figure~\ref{fig:setup}(b) as a function of mass concentration for the nanofibre suspensions. The crowding numbers for the nanorod suspensions are not plotted as the crowding number definition is only applicable to slender rods~\cite{Kerekes2006}. Indeed, despite a higher volume fraction of nanorods than nanofibres, the crowding numbers of the nanorods remain around unity due to their small aspect ratio. In Figure~\ref{fig:setup}(b), all the nanofibre suspensions exhibit a mean crowding number higher than the percolation threshold ($\langle N_c \rangle=16$) showing that the interactions between the nanofibres are important for all cases. At the maximum concentration, most of the nanofibre suspensions (except for CNF-$1360$) are above the rigidity threshold ($\langle N_c \rangle=60$). This clearly emphasizes that the nanofibres are highly entangled and constrained in a three dimensional network. Note that the mean crowding number does not represent the proper entanglement for all nanofibres. Indeed, the entanglement of a nanofibre originates from the number of contact points with its neighbours. Therefore, the shorter the nanofibres, the smaller the number of contacts and less the entanglement. The entanglement length $L_*$ defined in the Supporting Information represents the threshold between small nanofibres in the dilute regime (not entangled) and the larger ones in the semi-concentrated regime (entangled). For the nanofibre suspensions used in this study, $L_*$ is always smaller than $125$~nm, showing that all nanofibres are in the semi-dilute regime (see Figure~\ref{fig:setup}(a)).

A typical birefringence signal before the stop of the flow is shown in Figure~\ref{fig:setup}(c), with the flow going from left to right. Before the focusing point ($x/h<0$), the maximum birefringence is located on the walls, due to alignment of the nanoparticles with shear~\cite{Jeffery1922}. In the focusing region ($0<x/h<2$), the particles are aligned by the extensional flow and the birefringence reaches its maximum value at around $x/h=1.25$. After the focusing region ($x/h>2$), the thread attains its final shape with no further mechanisms causing alignment. Thus, the particles are relaxing towards isotropy due to rotational diffusion, while being advected by the flow.

\subsection{Orientational dynamics in polydisperse networks}

\begin{figure*}[b!]
\begin{center}
  \includegraphics[width=1\linewidth,clip=]{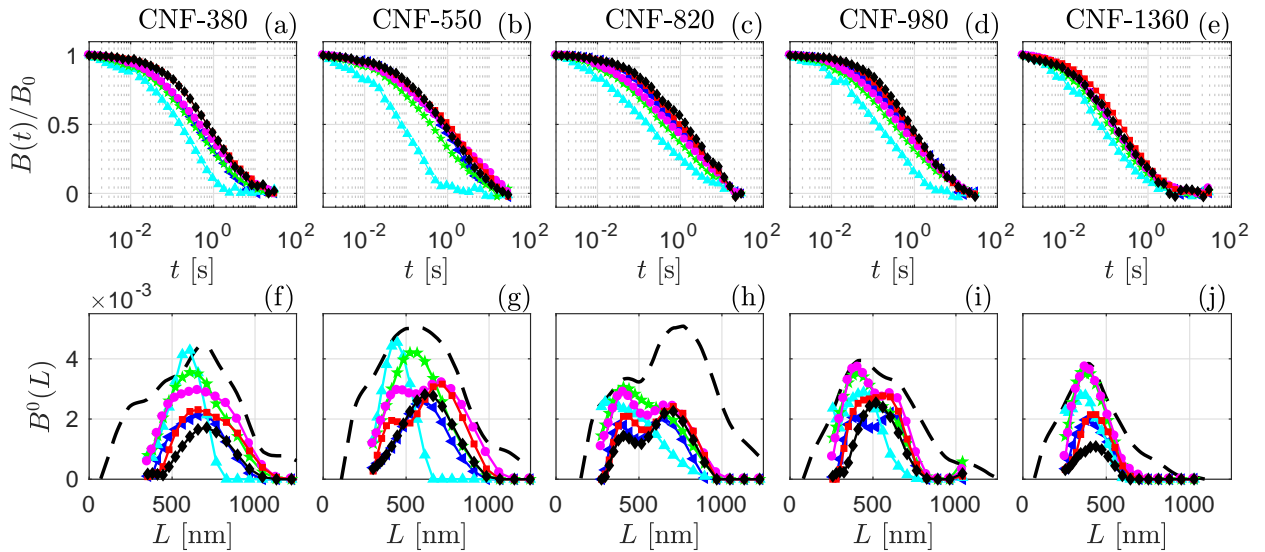}
   \caption{Orientational dynamics as a function of length distribution. Normalized birefringence decays~(top) and contributions~$B^0(L)$~(bottom) for all the nanofibre suspensions at $C_m=3$~g/l. The length distributions of the suspensions are represented by the dashed black line in the bottom panels. The colours correspond to different locations in the channel, as shown in Figure~\ref{fig:setup}(c): $x/h=1$ (blue), $x/h=0.75$ (light blue), $x/h=1.25$ (green), $x/h=1.75$ (magenta), $x/h=4.75$ (red) and $x/h=8$ (black).}\label{fig:length_variation}
\end{center}
\end{figure*}

Figure~\ref{fig:length_variation} shows normalized birefringence decays $B(t)/B_0$ and the corresponding contributions~$B^0(L)$ after inverse Laplace transform for the five nanofibre suspensions at $C_m=3$~g/l. The data represented here originate from the specific locations marked with the coloured squares in Figure~\ref{fig:setup}(c). The length distributions obtained with TEM images are plotted as a dashed black line in the bottom panels of Figure~\ref{fig:length_variation}. The general trend for the orientational dynamics is found to be size-dependent~\cite{FlowStopLetter2018} and similar for all suspensions. At the focusing point (light blue triangles), short nanofibres are first aligned by the flow and the birefringence decays therefore exhibit short time scales. Further downstream, the decays show longer time scales, highlighting that long nanofibres have been aligned by the flow and that the short ones have already started to dealign. Note that the scale separation between the short and long aligned nanofibres depends on the width of the length distribution. For the narrow length distribution of CNF-$1360$, the scale difference is small and most of the decays collapse while for a wider distribution as CNF-$550$, the scale difference is large and the decays are very different depending on the downstream location. 

The $\beta$ parameter in eq~\ref{eq:polydisperse} is estimated from the comparison with the TEM length distribution. When the peculiar details of the length distributions are well fitted by the contributions $B^0(L)$, like for example CNF-$1360$ which is relatively narrow or CNF-$980$ which presents a double shoulder, the determination of $\beta$ is straightforward. However, for the wide length distributions of CNF-$380$, $550$ and $820$, the estimation of $\beta$ is more difficult because the flow is not able to align all the nanofibres in the length distribution~\cite{FlowStopLetter2018}. Nevertheless, it is considered that a majority of the nanofibres are aligned by the flow and that $\beta$ is independent of mass concentration. This allows to average the values obtained for the same nanofibre sample at three different concentrations and provides error bars. The determination of $\beta$ is dependent on the diameter $d$ of the samples used, hidden in $L_*$. As the entanglement length $L_*$ is to the power of $4$ in eq~\ref{eq:polydisperse}, a small error in $d$ leads to a significant change in $\beta$. For example, the $\beta$ value found here for CNF-$380$ is around $750$ while the value reported previously~\cite{FlowStopLetter2018} using the same data was twice this value. This discrepancy is due to the fact that the diameter used here, $d=3.62$~nm, has been taken from SAXS data~\cite{Gengetal2018} while the diameter used previously~\cite{FlowStopLetter2018}, $d=3$~nm, has been taken from AFM measurements.

\begin{figure}[t!]
\begin{center}
  \includegraphics[width=0.5\linewidth,clip=]{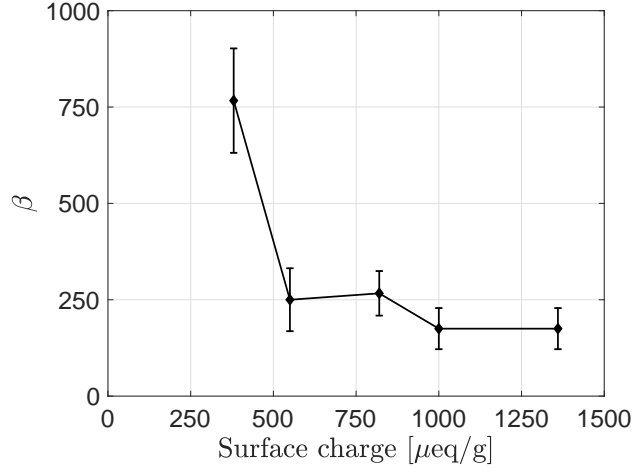}
   \caption{$\beta$ parameter as a function of the surface charges of the different nanofibre samples.}\label{fig:beta}
\end{center}
\end{figure}

Figure~\ref{fig:beta} shows the $\beta$ parameter in eq~\ref{eq:polydisperse} as a function of the nanofibre surface charges, obtained by comparing the length distribution with the contributions $B^0(L)$ in Figure~\ref{fig:length_variation}. The $\beta$ parameters found here are in the range $100-1000$, i.e. slightly lower than the values reported by other experiments~\cite{Rogersetal2005,KeepPecora1985,Chowetal1985,DoiEdwards1986}. They show a decreasing trend when the surface charges increase and seems to reach a plateau around $200$. This trend can be understood using several arguments. As surface charges increase, the repulsion between the rods is higher and could influence the rotational diffusion. However, as surface charges are highly correlated with length distributions, this trend can also be associated with length effects. Indeed, the value of $\beta$ is known to be strongly dependent on the caging of the nanorods~\cite{KeepPecora1985,Odelletal1983,JainCohen1981,TracyPecora1992,Teraokaetal1985,Chowetal1985,Rogersetal2005}, i.e. the entanglement in the network. As different length distributions cause different networks, this could lead to variations in $\beta$ values between the samples. 

\subsection{Characteristic length scales aligned by the flow}

\begin{figure}[b!]
\begin{center}
  \includegraphics[width=0.5\linewidth,clip=]{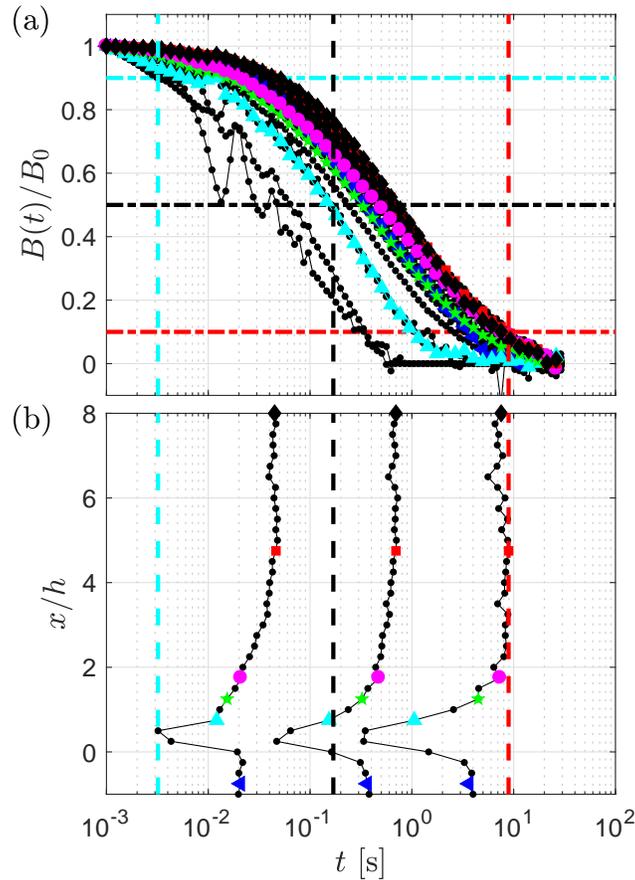}
   \caption{Extraction of reduced quantities. (a) Normalized birefringence decays for the CNF$-380$ suspension and for all positions. The horizontal dashed-dotted lines indicate $10\%$~(light blue), $50\%$~(black) and $90\%$~(red) of the decay. (b) Relevant time scales as a function of position for each decay: $t_{10\%}$~(left), $t_{50\%}$~(centre) and $t_{90\%}$~(right). The vertical dashed lines indicates the reduced time scales extracted from the plot: $t_{min}=\min(t_{10\%})$~(light blue), $t_{max}=\max(t_{90\%})$~(red) and $\bar{t}=(t_{min}t_{max})^{1/2}$~(black). The coloured points in panels (a) and (b) correspond to the decays obtained in the coloured squares in Figure~\ref{fig:setup}(c) and are here to help the eyes.}\label{fig:reduced_quantities}
\end{center}
\end{figure}

To quantify quickly and compare quantitatively the birefringence decays and the different length fractions aligned by the flow at all positions and for all suspensions, three characteristic time scales, $\bar{t}$, $t_{min}$ and $t_{max}$, have been defined as relevant reduced quantities. The determination of these reduced quantities is demonstrated in Figure~\ref{fig:reduced_quantities} with the CNF-380 suspension at $C_m=3$~g/l. Figure~\ref{fig:reduced_quantities}(a) shows the birefringence decays at the $37$ positions marked by the squares along the centreline in Figure~\ref{fig:setup}(c). For the specific positions highlighted by the coloured squares, the symbols of the corresponding decays are also coloured and slightly larger in Figure~\ref{fig:reduced_quantities}(a) while the symbols for the decays at other locations are simple black dots. All the decays are normalised by the initial birefringence $B_0$ so all curves start from $1$ for small time scales and end at $0$ for large time scales. The three horizontal dashed dotted lines indicate $10\%$~(light blue), $50\%$~(black) and $90\%$~(red) of the decay. Figure~\ref{fig:reduced_quantities}(b) shows the different times, named $t_{10\%}$, $t_{50\%}$ and $t_{90\%}$, where the decays are crossing these three lines, as a function of the downstream location of the decays $x/h$. Again, the symbols used are black dots except for specific locations shown in colours. These three time series have different orders of magnitude but exhibit a very similar behaviour as a function of the downstream location. Before focusing ($x/h<0$), the values are almost constant. Then, around the focusing point ($0<x/h<2$), the time scales drop suddenly down to a minimum before increasing towards a final value. They finally reached a plateau further downstream ($x/h>5$). These short time scales around the focusing point are associated with the alignment of the short nanofibres by the flow while the larger time scales further downstream are the signature of the long nanofibres that remain aligned far from the focusing point~\cite{FlowStopLetter2018}. Therefore, the shortest time scale in Figure~\ref{fig:reduced_quantities}(b), defined as $t_{min}=\min(t_{10\%})$ and shown with a vertical dashed light blue line in both panels, corresponds to the shortest nanofibres that are aligned by the flow in the suspension. Similarly, the longest time scale in Figure~\ref{fig:reduced_quantities}(b), defined as $t_{max}=\max(t_{90\%})$ and shown with a vertical dashed red line in both panels, is associated to the longest nanofibres aligned by the flow in the system. In the example shown in Figure~\ref{fig:reduced_quantities}, these two time scales are separated by more than three orders of magnitude. The geometric mean of these two quantities, $\bar{t}=(t_{min}t_{max})^{1/2}$, therefore gives the mean time scale of the decays. The time scale $\bar{t}$ is plotted as a vertical dashed black line in both panels of Figure~\ref{fig:reduced_quantities}. It corresponds well to the geometric mean of the minimal and maximal values of the time scales $t_{50\%}$, showing that this time scale is relevant as a mean time scale. 

\begin{figure}[t!]
\begin{center}
  \includegraphics[width=0.5\linewidth,clip=]{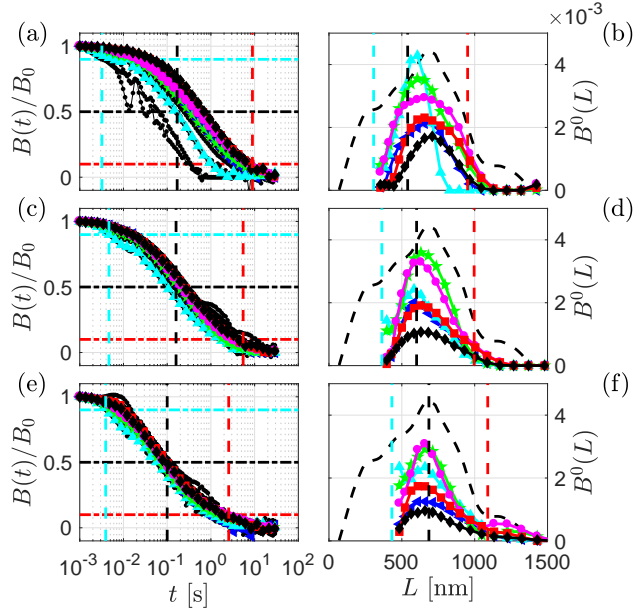}
   \caption{Comparing inverse Laplace transform and reduced quantities. Normalized birefringence decays~(a) (respectively~(c) and~(e)) and contributions~$B^0(L)$~(b) (resp.~(d) and~(f)) for the CNF-$380$ suspension at $C_m=3$~g/l (resp. $C_m=2$~g/l and $C_m=1$~g/l). The length distribution of the suspension is represented by the dashed black line in panels (b),~(d) and~(f). The vertical dashed lines show $t_{min}$ or $L_{min}$ (light blue), $\bar{t}$ or $\bar{L}$ (black) and $t_{max}$ or $L_{max}$ (red).}\label{fig:concentration_variation}
\end{center}
\end{figure}

The typical length scales aligned by the flow, $\bar{L}$, $L_{min}$ and $L_{max}$, are obtained by converting the three characteristic time scales, $\bar{t}$, $t_{min}$ and $t_{max}$, into length scales using the polydisperse diffusion model~\cite{MarrucciGrizzuti1983,MarrucciGrizzuti1984} (see eq~\ref{eq:polydisperse}) and the $\beta$ values obtained previously. Figure~\ref{fig:concentration_variation} shows on the left the normalized birefringence decays at all positions for the three CNF-$380$ suspensions at different concentrations.  The corresponding contributions $B^0(L)$ obtained after inverse Laplace transform of only six decays at specific locations are shown in Figure~\ref{fig:concentration_variation} on the right, together with vertical lines indicating the length scales $\bar{L}$, $L_{min}$ and $L_{max}$. For the different concentrations, $L_{min}$ and $L_{max}$ demarcate very clearly the contributions~$B^0(L)$ while $\bar{L}$ fits well the most aligned length in the suspension. Similar results are also obtained for all the other suspensions. This shows that the reduced quantities capture well the main characteristics of the contributions~$B^0(L)$ without performing inverse Laplace transforms and are therefore relevant to compare all suspensions investigated in this paper. Note that the general shape of the contributions~$B^0(L)$ is modified as the concentration decreases: from a symmetric shape at $C_m=3$~g/l, it becomes more and more asymmetric with a large tail towards long nanofibres at lower concentrations. Consequently, the maximal aligned length $L_{max}$ increases as the concentration goes down as shown by the red dashed lines in Figure~\ref{fig:concentration_variation}. Moreover, the maximum of the contributions~$B^0(L)$ shifts towards longer lengths as a function of the downstream location~$x/h$ at $C_m=3$~g/l while it remains approximately the same for lower concentration. This is consistent with the fact that the decays $B(t)/B_0$ are more collapsed at low concentrations and shows that concentration affects greatly the length fractions aligned by the flow.

\begin{figure}[t!]
\begin{center}
  \includegraphics[width=0.5\linewidth,clip=]{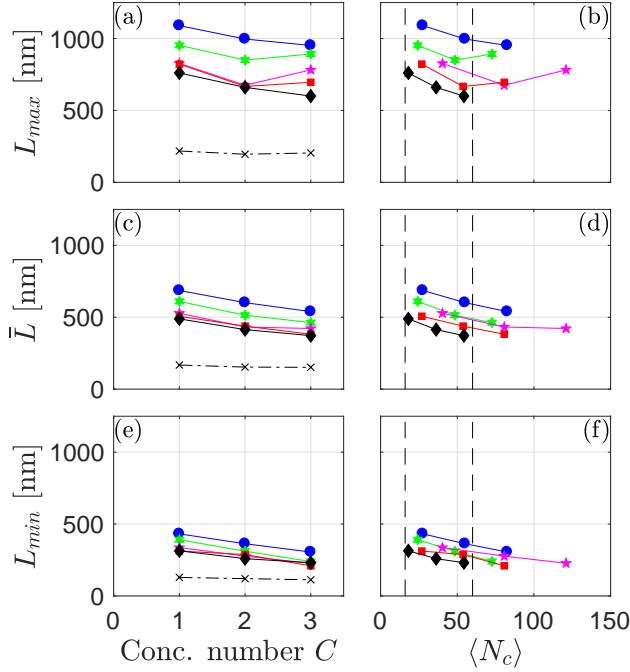}
   \caption{Flow-aligned length scales as a function of entanglement. Characteristic length scales as a function of concentration number (a, c and e) or as a function of the crowding number (b, d and f). Each colour corresponds to one nanofibre suspension, as in Figures~\ref{fig:setup}(a) and (b). The dashed dotted lines with the crosses corresponds to the nanorod suspensions. $L_{max}$ is shown in panels (a) and (b), $\bar{L}$ in panels (c) and (d) while $L_{min}$ is represented in panels (e) and (f). The dashed vertical lines in panels (b), (d) and (f) indicate the two entanglement thresholds $\langle N_c \rangle=16$ and $\langle N_c \rangle=60$.}\label{fig:length_scales}
\end{center}
\end{figure}

The three characteristic length scales of all suspensions are shown as a function of concentration number and crowding factor in Figure~\ref{fig:length_scales}. The concentration number $C$ is defined as $C_1=1$~g/l, $C_2=2$~g/l, and $C_3=3$~g/l for the nanofibres and as $C_1=11$~g/l, $C_2=20$~g/l and $C_3=41$~g/l for the nanorods, in order to present all the samples in Figure~\ref{fig:length_scales}. For all nanofibre suspensions (coloured symbols and solid lines), the three length scales generally decrease when the concentration or crowding number increases, showing that it is more difficult to align longer nanofibres at large concentration or entanglement. However, these length scales for the nanorods shown in Figures~\ref{fig:length_scales}(a), (c) and (e) (dashed dotted lines and crosses) are independent of the mass concentration. This highlights the importance of the network in polydisperse suspensions, where flow-aligned length scales are selected depending on entanglement, with a subtle interplay between Brownian rotational diffusion and flow-alignment.

\subsection{Transition from mobility to rigidity}

Thoroughly describing the \textit{in-situ} orientational mechanisms within polydisperse nanofibre networks requires the comparison of three different quantities for each nanofibre: the rotational diffusion coefficient $D_r$~(see eq~\ref{eq:polydisperse}), the alignment rate~$\dot{\varepsilon}$ and the typical time scale~$\tau_a$ during which the alignments mechanisms are applied. For a single nanofibre of high aspect ratio in a viscous flow, the alignment rate is independent of the nanofibre length~\cite{Jeffery1922}. However, in the case of a network, it is considerably reduced by the entanglement of the fibre and is therefore expected to depend on the length of the fibre. The time scale $\tau_a$ here is the time spent by the nanofibres in the extensional flow at the focusing point, i.e smaller than a second, but can be generalized for other alignment mechanisms.

In polydisperse suspensions, small nanofibres are poorly entangled but cannot be aligned because Brownian diffusion dominates flow-alignment ($D_r \gg \dot{\varepsilon}$). As the entanglement increases with nanofibre length, there is a first alignment threshold where diffusion and flow-alignment equilibrates ($D_r \approx \dot{\varepsilon}$). This threshold gives the typical size $L_{min}$ of the smallest nanofibres aligned by the flow. For nanofibres longer than $L_{min}$, i.e. for higher entanglement, flow-alignment overcomes Brownian diffusion ($D_r < \dot{\varepsilon}$) and these nanofibres become aligned until a second entanglement threshold is reached. At this threshold, the entanglement is more important such that diffusion becomes negligible but the flow is not able to align the nanofibres anymore, i.e. $\dot{\varepsilon} \tau_a \approx 1$. Indeed, the alignment rate $\dot{\varepsilon}$ is reduced by entanglement and therefore too weak to align the nanofibres during the time scale $\tau_a$. This second entanglement threshold gives the length of the longest nanofibres aligned by the flow $L_{max}$. For some nanofibre suspensions, note that $L_{max}$ is enhanced at $C_m=3$~g/l compared to the values at $C_m=2$~g/l. This appears for the CNF-$550$, $820$ and $980$ suspensions that have reached very large crowding numbers above the rigidity threshold $N_c=60$. At such large crowding numbers, the system becomes close to rigidity with a large number of contact points between nanofibres. Consequently, as the whole network is under hydrodynamic extensional forces, a slightly rigid network may improve flow-alignment of individual nanofibres instead of reducing it and therefore enhances the maximal length of the aligned nanofibres.

\begin{figure*}[b!]
\begin{center}
\includegraphics[width=0.9\linewidth,clip=]{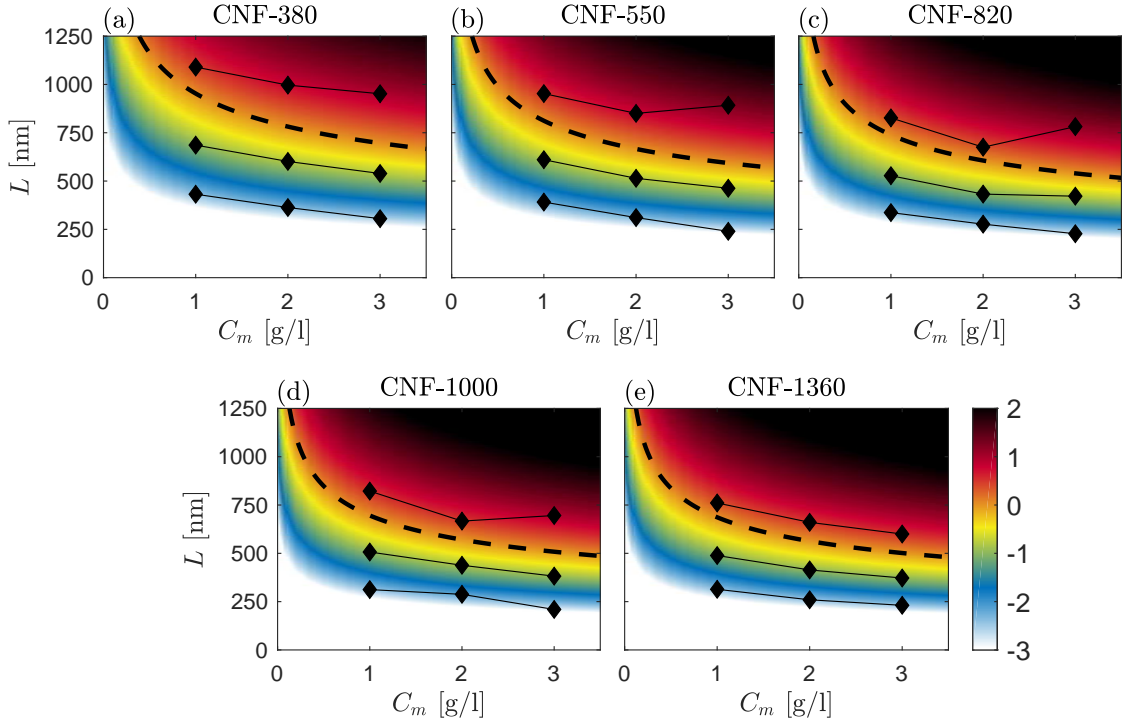}
   \caption{Diffusion time scales of flow-aligned length scales. Logarithm of diffusion time scale $t_D$ as a function of the length $L$ and the mass concentration $C_m$ for the nanofibre suspensions. The black dashed lines indicate the gelling time scale $\tau \approx1$~s. The symbols show the typical length scales aligned by the flow: $L_{max}$ (top), $\bar{L}$ (center) and $L_{min}$ (bottom).}\label{fig:diffusion_map}
\end{center}
\end{figure*}

The physical descriptions of these two entanglement thresholds are further confirmed by plotting the aligned length scales $\bar{L}$, $L_{min}$ and $L_{max}$ in Figure~\ref{fig:diffusion_map} as a function of mass concentration with black diamonds. The colormap shows the logarithm of diffusion time scale $t_D$ given by the polydisperse diffusion model~\cite{MarrucciGrizzuti1983,MarrucciGrizzuti1984} (see eq~\ref{eq:polydisperse}), mainly depending on the length of the nanofibres and on the concentration. $L_{min}$ and $\bar{L}$ as a function of mass concentration~$C_m$ follow diffusion isochrones in Figure~\ref{fig:diffusion_map} and therefore correspond to invariant diffusion time scales, i.e. to the same entanglement thresholds independent of the mass concentration. Flow-alignment for these length scales is thus balanced by Brownian rotational diffusion and can be described similarly through the individual entanglement of the nanofibres, namely the number of contacts or neighbours. This illustrates the strong interplay between flow-alignment and Brownian diffusion in a fibre network governed by entanglement. However, the maximal aligned length $L_{max}$ is not associated to a single diffusion time scale in Figure~\ref{fig:diffusion_map} as flow-alignment for long nanofibres is not balanced by diffusion but is limited by the entanglement and the time scale $\tau_a$. Therefore, the transitions of rotational mobility and rigidity in polydisperse nanofibre networks are governed by two dimensionless numbers, $D_r / \dot{\varepsilon}$ and $\dot{\varepsilon} \tau_a $ respectively.

\subsection{Application to high-performance nanostructured materials}

Elevated mechanical performances in nanostructured materials can be obtained by maximizing the physical contacts between the nanoparticles through the orientation at the nanoscale. This allows an improved stress transfer between the nanoparticles and provides the ability to utilize the nanoscale mechanical properties in the macroscale materials, which is considered as one of the largest challenges in material science till date.~\cite{Kiriyaetal2012,NatureCom2014,Mittaletal2018}. The fabrication of these hierarchical superstructures through self-assembly after flow-induced alignment is generally achieved by the ionic cross-linking of the nanoparticles or by tuning the nanoparticle concentration in the suspensions~\cite{Rammenseeetal2008,Kangetal2012,NatureCom2014,Haynletal2016,Kamadaetal2017,Mittaletal2018,Jahnetal2004,Schabasetal2008}. In the extensional flow-based geometries of similar dimensions, the typical time $\tau$ to complete the transition to colloidal glassy-state and to lock the nanoparticles in the aligned state within the thread is of the order of a second~\cite{NatureCom2014,Mittaletal2018}. This time is in competition with the typical rotational diffusion time scale $t_D$, highly dependent on the length of the nanofibres. The comparison of these two time scales is performed in Figure~\ref{fig:diffusion_map} for all nanofibre suspensions, where the dashed black line shows the transition time scale $\tau=1$~s. Both $\bar{L}$ and $L_{min}$ are below this line, for all concentrations, meaning that nanofibres of these lengths have already dealigned significantly before the transition occurs. Consequently, only the very long nanofibres having a diffusion time scale $t_D$ higher than the transition time~$\tau$ remain aligned before the transition occurs. Therefore, the suspensions having aligned nanofibres the farthest above the dashed lines in Figure~\ref{fig:diffusion_map}, i.e. CNF-$380$, $550$ and $820$ at $C_m=3$~g/l, would lead to the macrostructures with the highest degree of orientation and the best mechanical properties as demonstrated by Mittal \latin{et al.}~\cite{Mittaletal2018}.

To maximize nanofibre alignment during the material fabrication, it is therefore necessary to work with long and partially entangled nanofibre systems, i.e. with broad length distributions and concentrated suspensions slightly above the rigidity threshold. Indeed, long nanofibres own a dealignment time scale larger than to the transition time and large concentrations lead to high entanglement and thus maximize the flow-induced alignment. Large concentrations are however limited by the rigid volume-spanning arrested states, where the suspension becomes like a gel due to entanglement. In such gel-like suspension, flow alignment is completely restricted. Hence, it is highly important to optimize the nanoparticle systems based on the dimensions and concentrations to obtain the best possible mechanical properties at the macroscale.

\section{Conclusions}

To conclude, we have demonstrated that tracking the rotational Brownian motion of nanorods with birefringence expands the capabilities to probe \textit{in-situ} the orientational and network dynamics of flowing suspensions at the nanoscale. Our results reveal that despite a similar trend in the orientational dynamics, the characteristic length scales aligned by the flow are strongly dependent on the nanofibre length distributions and on the nanoscale entanglement within the suspensions. The selection of these aligned length scales emphasizes the intense coupling between Brownian diffusion and flow-induced alignment at play in entangled nanofibre networks. The knowledge from this study can be used to optimize the mechanical performance of nanostructured materials fabricated \textit{via} flow-based approaches, that are typically dependent on the nanoparticle length distributions and characteristic time scales of diffusion. Our approach promises an easy access to nanoscale mechanisms, dynamics and structure for a variety of biophysical and soft material systems that are currently accessible only to the state of art X-rays and neutron scattering methods and are challenging to characterize \textit{in-situ}. In the examples provided herein, the nanoparticle building blocks (nanocelluloses) differ only in size, but this methodology also should be extendable to other building blocks that differ in chemical compositions (i.e. amyloids, chitin, metal nanoparticles etc...).

\begin{acknowledgement}

The Wallenberg Wood Science Center at KTH is acknowledged for providing the financial assistance. The authors are thankful to Dr. Assya Boujemaoui for providing the nanorod suspension and to Dr. Michaela Salajkova for assistance with the TEM images. Dr. Tomas Ros{\'e}n is acknowledged for experimental assistance and helpful discussions.

\end{acknowledgement}

\section{Notes}
The authors declare no competing financial interest.

\begin{suppinfo}

Details about the preparation of the nanocellulose suspensions. 
TEM and AFM images for nanocellulose characterization in dry state (Figures S$1$ and S$2$).
Description of the Inverse Laplace Transform procedure.

\end{suppinfo}

\bibliography{flow_stop}

\end{document}